# A general designing approach for polarization-independent photonic on-chip systems

YATING WU[1], XIAOYAN LIU[1], TAO CHU[1,*]

[1] College of Information Science and Electronic Engineering, Zhejiang University, 38 Zheda Road, Hangzhou 310027, China
*Corresponding author: chutao@zju.edu.cn

*Abstract:*

*Due to that the polarization states in optical fibers change randomly during transmission, polarization-independent (PID) devices are demanded to receive lights with arbitrary polarization states. Compared with their orthogonal polarization states, the optical profiles of various modes of the same polarization are similar, and their directions of the main electric field are same. Therefore, it's much easier to design PID devices using multi-modes of one polarization state instead of orthogonal polarizations. This paper firstly presents a scalable method to achieve PID devices and systems by transforming the light orthogonal polarization states into one polarization with different modes. Taking thermo-optical switches as an example, the PDL of the fabricated 2 × 2 switch cell was about 0.8 dB at 1300-1360 nm, and its extinction ratio (ER) was larger than 19 dB for both polarizations. A 4 × 4 switch was also demonstrated and its function of dual polarization switching was implemented, whose ER was larger than 11 dB for both polarizations. This method has potential in constructing PID on-chip systems for optical communications, optical interconnections, etc.*

Owing to the advantages of CMOS compatibility, low latency, low power consumption, and large bandwidth, silicon-based integrated optoelectronic devices have shown a wide range of applications in optical communication, optical interconnection, and optical computing[1-6]. When a silicon-based photonic device or system works without oh-chip integrated light sources, it receives the light from off-chip lasers with ordinary single-mode optical fibers, whose polarization state is usually not single linear polarization owing to the inevitable fiber bending and entanglement. On the other hand, due to the need of high-density integration, the waveguide section size is usually restricted to the propagation limit of single mode light, which determines its high sensitivity to polarization. To adapt to the randomness of the polarizations of the light from optical fiber and receive the total optical signal, photonic devices and systems that are not sensitive to polarization states are highly required.

There are some relevant researches reported. One method is to introduce complex waveguide structures such as subwavelength grating[7,8] to weakened birefringence under single-mode waveguide conditions, which requires high design capabilities and the delicate fabrication process. The second method is to adopt the waveguide with a square cross-section[9] to reduce its waveguide birefringence, but it is difficult to fabricate a perfect square cross-section due to the existence of process error and interface stress, besides that the upper and down claddings bring more complicated strain birefringent effects.

The third method is to add an active polarization control system[10-14], which converts the random polarization state in the fiber to a single linear polarization before entering the functional device. Its mature scheme is to separate and rotate the orthogonal polarization state into one polarization firstly, and then achieve beam closure by adjusting the phase of the two beams. However, the active control operation increases the power consumption, complexity of packaging, and circuit control. Most importantly, it limits the work for multiple wavelength light signals with different polarizations.

The fourth method is the polarization diversity scheme[15-18], which splits random orthogonal polarization state into two outputs with the same polarization using a polarization splitter and polarization rotator first, and then the light is processed by two sets of functional devices, and finally combined together after polarization rotation and combination for output However, with the expansion of the scale of functional devices, the scheme almost doubles the chip area and power consumption, and increases the difficulty of circuit control system and packaging. Based on this, the non-duplicated polarization diversity was adopted to avoid the two sets of systems[19,20]. However, only specific and limited networks can meet the non-duplicated requirements.

Here we propose a scalable method to implement polarization-independent (PID) functional devices with the help of multi optical modes. By converting the input random light polarizations with fundamental TE0 and TM0 mode components into different modes of the same polarization, the functional device does not need to meet the polarization insensitivity but rather mode insensitivity. Since the modulation efficiency and dispersion curves of different modes of the same polarization are closer, which gets rid of the limitation of process level was mitigated, and the design difficulty was simplified. By using a mode converter at the input of the functional device or system to convert the input TM0 mode to TE1 mode while keeping the TE0 mode passing through, the PID requirement is transformed into a simpler multi-mode requirement. The TE1 mode light is converted back to TM0 mode light and combined back with processed TE0 light at the output, so as to achieve the polarization independence.

## 1. Conception

Due to the difference in the electric field of different polarizations and light field distribution in the waveguide, electrical modulation efficiencies for TE and TM are different. The overlap area between the optical profile with the thermal field is also different, so the thermo-optic modulation efficiencies of TE and TM lights are different. However, for different modes of the same polarization, the difference can be easily reduced because the direction of the main electric field and the profile distribution are highly similar. In order to achieve PID functional devices, our main idea is shown in Figure 1. The TM0 was converted into TE1 through the input mode convertor (MC), while TE0 wasn't changed. The functional devices process the TE1 and TE0 signals simultaneously. After processing, the TE1 signal is converted back to the TM0 signal and combined with processed TE0 signal together for output. Thus, the input signal with TM0 and TE0 mode components are processed simultaneously with the same one set of devices or systems and achieve the polarization independence. This approach also shows the potential to be applied to the design of on-chip photonic systems with active and passive components such as modulators, cross-waveguides, microrings, etc.

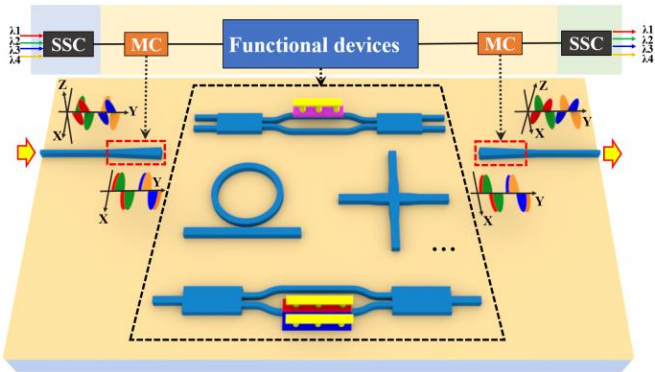

Figure 1. PID system conception

Here a thermo-optical PID 2 × 2 optical switch cell is taken as an example, which consists of two mode converters at the input and output ends, and a mode-independent Mach–Zehnder interferometer (MZI). The three-dimensional schematic diagram is showed in Figure 2.

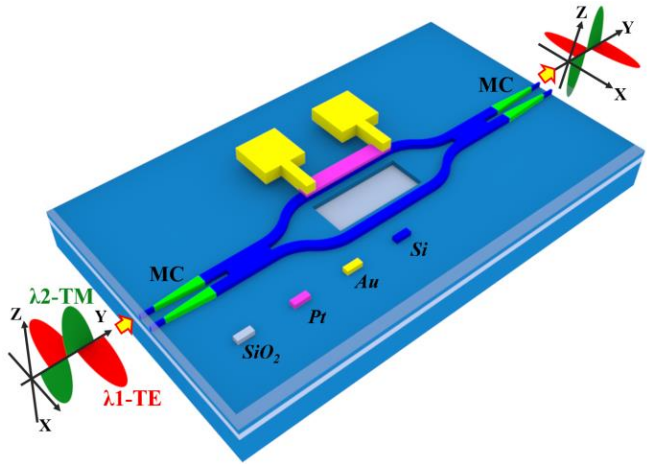

Figure 2. Schematic diagram of a PID optical switch

In this 2 × 2 switch cell, the input light TM0 component was converted into TE1 mode through the input mode convertor (MC), while TE0 component wasn't changed. These two modes of TE0 and TE1 were split into two light beams with the same amplitude but phase difference of π/2 with the input multimode interference (MMI) coupler. They were merged by the output multimode interference (MMI) coupler when no voltage was applied and final output from the switch cross port. When the phase difference of the MZI phase arms induced by the heater was π, the two light beams from the input MMI interfered with each other and output from the bar port to realize its PID switching operation.

## 2. Simulation

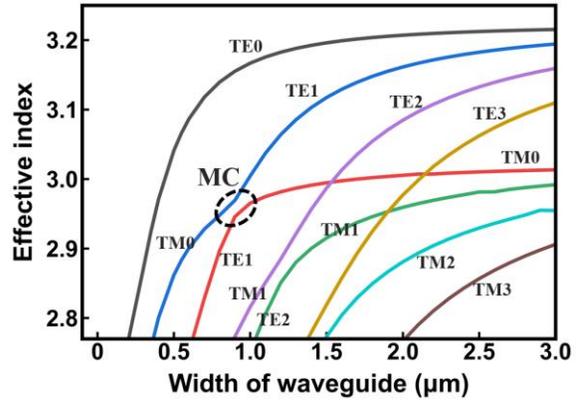

Figure 3. The curve of the mode refractive index vs. the width of the waveguide

The devices demonstrated here were all designed based on a SOI substrate with a 340-nm-thick top silicon layer and a 2-μm-thick BOX layer. The rib waveguide of 190-nm etching depth was adopted in all devices.

*A. MC*

The curve of effective index vs. waveguide width was simulated through Lumerical FDE and shown in Figure 3. The mode conversion of TM0-TE1 happened when the width of the waveguide was about 0.87 μm. Therefore, the width of the mode converter (MC) was chosen from 0.8 to 0.95 μm, while the length of MC (MCL) was swept from 50 μm to 400 μm. Mode conversion efficiency is larger than 97% when L is longer than 150 μm. In order to increase its process tolerances and improve the conversion efficiency as much as possible, an MCL of 400 μm was chosen. The light propagation profile at 1310 nm was shown in Figure 4(a)(b) when TE0 and TE1 were input separately. Figure 4(c)(d) illustrated that the loss for single MC of TE0 and TE1 mode light was smaller than 0.05 dB and 0.1dB and the polarization-dependent loss (PDL) was smaller than 0.05 dB in the total O band. The TM0 light crosstalk for TE1 light input was less than 20 dB.

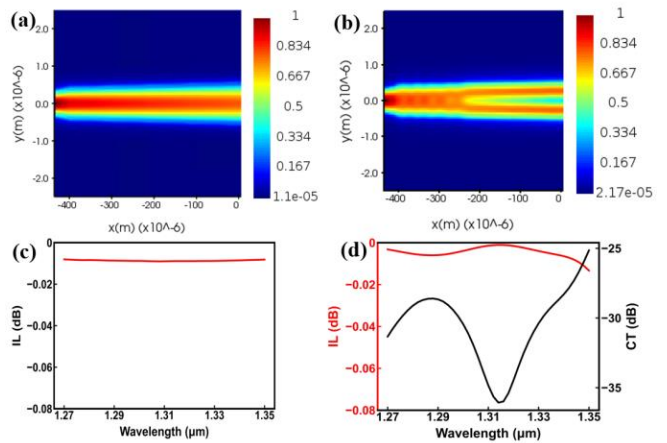

Figure 4. profile of MC when (a) TE0; (b) TE1 input @1310nm;(c) IL and crosstalk vs. the wavelength for (c)TE0; (d) TE1.

*B. MMI*

For the case of general interference, input light was evenly split into two output ports when the length of the MMI multimode area is about 1.5 Lπ[21], namely a 3 dB splitter. Because the length of the multimode area was quadratic to its equivalent width, the width of 5 μm was chosen to trade off the distance of the output ports and the size of the device. The length of the multimode area was chosen to be 112 μm after FDTD sweeping. The optical profile at 1310 nm and wavelength-dependent curves were shown in Figure 5. In Figure 5(c), the imbalance was smaller than 0.08 dB from 1.29 to 1.33 μm for TE0 and TE1 modes, and the insertion loss was smaller than 0.05 dB for TE0 and 0.2 dB for TE1 modes so that the mode-dependent loss was smaller than 0.15 dB. In Figure 5(d), the phase difference between the two output ports ranges from +0.4° to -0.7° near ideal 90° for TE0 and TE1 modes. Therefore, for these two mode lights, the MZI composed of MMI can achieve the same switching function under the same phase shift.

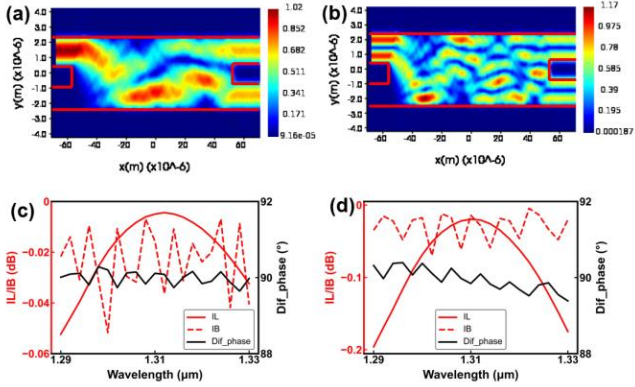

Figure 5. profile of mode-independent MMI when (a)TE0; (b)TE1 input @1310nm;(c) IL, IB, and (d)phase difference between output ports vs. the wavelength for TE0 and TE1.

*C. phase-shifted arm*

The relationship of phase shift vs heat power was simulated through Lumerical Heat. Because of distinct optical profile distribution under the same heat profile and orthogonal principal electric field component, overlapping integrals between the optical profile and heat profile are different, so the modulation efficiency is different for TE0 and TM0. However, when the width of the waveguide increased away single-mode condition, overlapping integrals with heat and the heat modulation efficiency of the different mode light of the same polarization approach gradually, especially for the two lowest order modes, while overlapping integrals of orthogonal polarization were still different. Therefore, orthogonal polarizations (TE0 and TM0) transformed into multi-modes (TE0 and TE1) of the same polarization can achieve similar heat modulation efficiency. Here, the modulation arm width of 2.2 μm was chosen.

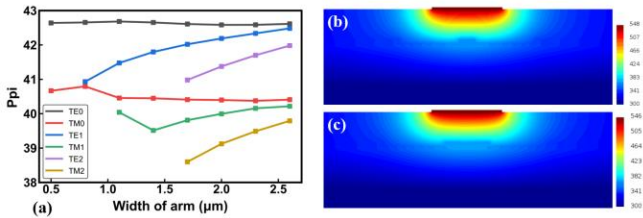

Figure 6. (a) Power consumption of phase shift $\pi$ for TE0, TM0, TE1, TE2, TM1 and TM2 under different widths of arm; profile with width of (b) 0.8 μm; (c)2.2 μm

## 3. Fabrication

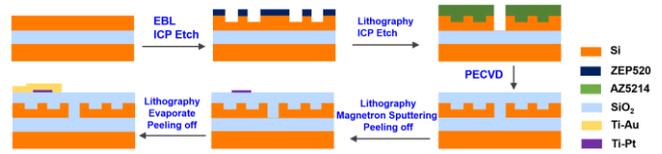

Figure 7. Fabrication flow.

After electron-beam lithography (EBL), the chip was etched by 190 nm through ICP etching. Then a 1.5-μm-thick silica cladding layer was deposited using PECVD. The pattern of the heater was formed by DWL, and then 100-nm-thick Pt was spurred and peeled off to form the thin-film heater. After the pattern of the Au electrode was formed by direct write laser (DWL) lithography, Au was evaporated and peeled off to form the electrode.

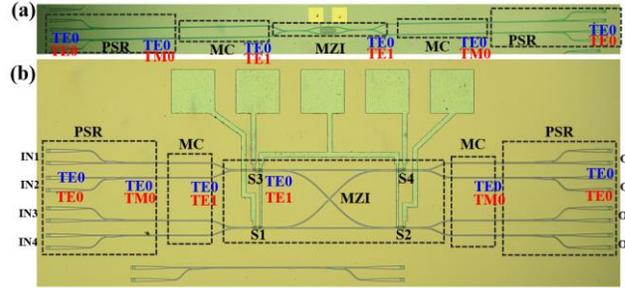

Figure 8. Micrograph view of the (a) 2 × 2 switch cell; (b) the 4 × 4 optical switches.

## 4. Testing

*A. Test system*

In measurements, the light was emitted from the tunable laser to the polarization control analyzer and then the three-ring polarization controller (PC). The light was coupled in and out of the chip with the grating couplers. Finally, the optical signal was transmitted to the optical power meter. The initial polarization state was calibrated to TE polarization by a polarizing beam-splitter prism and PC, which was changed to its orthogonal polarizations by a polarization control analyzer.

As for the electrical path, the electrical signals from the Precision voltage source were supplied to the electrode by the probe. The computer controls the voltage source and accepts the information from the optical meter.

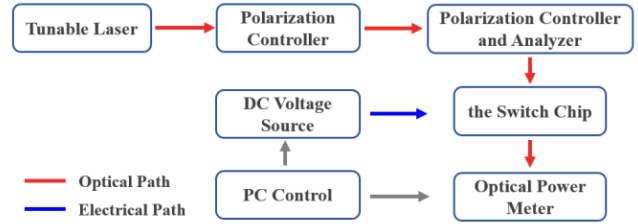

Figure 9. Test system.

*B. Test of MC*

The coupling loss of the grating coupler for TE0 mode was about 4 dB/facet. Figure 10(a) illustrated that in a TE1-assisted polarization splitter and polarization rotator (PSR) consisting of MC and mode multiplexer (mux) (inset in Figure 10(a)), TE0 was mainly output from the bar port, whose on-chip loss was about 0.5 dB in a range of 100 nm. TM0 was mainly output from the cross port, whose loss was about 1.5 dB. In addition, polarization ER was above 15 dB for both polarizations.

The function of MC was demonstrated by this PSR that TE0 does not transform and TM0 converts to TE1. Through cascade testing, loss of MC was measured at 0.2 dB for TM0 and 0.05 dB for TE0 mode light from 1270 to 1350 nm, shown in Figure 10(b).

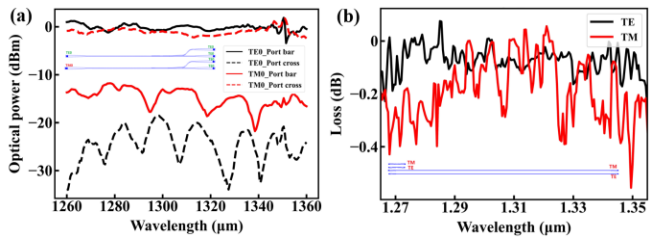

Figure 10. Wavelength-dependence test curves for (a) PSR with MC; (b) single MC loss.

*C. Test of switch cell*

A PSR can split orthogonal polarizations and rotate one polarization into the other. In a forward 1 × 2 PSR shown in Figure 10(a), TM0 was changed into TE0 at the cross port, while TE0 was not changed and output from the bar port of the PSR, Similarly, reverse 2 × 1 PSR can convert TE0 at the cross port into TM0, while maintaining TE0 at the bar port unchanged. Therefore, a 2 × 1 PSR at the input of the chip was used to form TE0 and TM0, and a 1 × 2 PSR at the output of the chip separated them, which were not parts of the PID switch and only for demonstrating the PID function. The loss of a pair of PSRs has been subtracted along with the gratings as the reference loss. As shown in Figure 8(a), the 2 × 2 switch cell consisted of input MC, MZI, and output MC. TM0 light from 2 × 1 PSR was converted into TE1 light by the input MC. After TE1 light was transmitted through the MZI, it was returned to TM0 light by the output MC. While for the TE0 light from the input PSR, it wasn't changed through the input MC, MZI, and the output MC. Finally, TE0 and TM0 from the output MC were collected to the bar port and the cross port of output PSR, respectively.

The 2 × 2 switch cell was measured under different polarizations. As shown in Figure 11(a), the power consumption vs. optical power(P-P) curve of the two polarizations almost coincide completely. When no voltage was applied, the switch was in the CROSS state. The loss on the chip was smaller than 0.5 dB for TE0 and TM0 light and the extinction ratio (ER) was larger than 19 dB at 1315nm wavelength. The PDL was about 0.3 dB. When the power applied was about 31.5 mW, the cell was switched to the BAR state. The loss on the chip was 0.4 dB and 0.7 dB separately for TE0 and TM0 and ER was both larger than 35 dB at 1315 nm. Therefore, a PID switch was achieved and proved in experiments. Figure 11(b) and Figure 11(c) illustrated wavelength-dependent properties in the CROSS and BAR states. Ranging from 1300nm to 1360 nm, the averaged PDL was smaller than 0.8 dB.

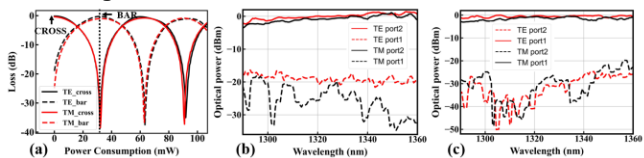

Figure 11. Voltage vs. (a) insertion loss (b) power consumption at the cross and bar ports for TM0 and TE0 input; Optical power of different ports at (c) CROSS state; (d) Bar state

*D. Test of 4 × 4 switches*

Based on the 2 × 2 switch cell, a 4 × 4 butterfly-typed switch was designed and fabricated. TE0 and TM0 formed by the input PSR were transmitted into input MC. TE0 was not changed, while TM0 was changed into TE1. Then TE0 and TE1 entered to switches. Through output MC, TE0 was not changed but TE1 was returned to TM0, and then TE0 and TM0 were transmitted and converted into the bar port and cross port by the output PSR, respectively.

Switch cells in the 4 × 4 switch were in all CROSS states for both polarization states, shown in Figure 12(a). The main optical path of in1-out4, in2-out2, in3-out3, in4-ou1 was achieved, and ERs were about 13dB. The driving voltages of four units are fixed to achieve all BAR states for both polarizations, respectively, shown in Table 1. The unit labeling is consistent with Figure 8(b). The main optical path of in1-out1, in2-out3, in3-out2, in4-ou4 was achieved and ERs were about 11 dB.

In the 4 × 4 switch, the waveguide width of 1.4 μm was selected for the waveguides except for the phase-shift arm. The simulation showed that the optical profile of TE1 mode was close to the waveguide sidewall. Due to the poor etching effect, the sidewall has periodic fluctuations, it brings large scattering losses to TE1 light, which can be mitigated by widening the waveguide or Optimized etching. In addition, since the maximum write field of EBL was 500 μm, the splicing error of the write field also had a serious impact on the transmission of TE1 light. These will be subsequently improved.

Table 1 Voltage (V) vs. switch cell at all BAR and all CROSS states

| State \ Switch | S1 | S2 | S3 | S4 |
|---|---|---|---|---|
| All CROSS states | 0 | 0 | 0 | 0 |
| All BAR states | 2.47 | 2.5 | 1.68 | 1.68 |

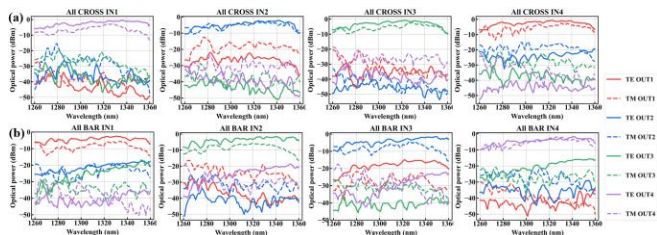

Figure 12. Test curves for 4×4 switch under (a) all CROSS states; and (b) all BAR states.

## 5. Conclusion

This paper proposed a method to achieve polarization insensitivity for on-chip photonic systems, by transforming the orthogonal polarization states of input light into a multi-order mode light of the same polarizations. Take thermo-optical switches as an example, the PDL of the fabricated switch cell was about 0.8 dB at 1310-1360 nm, and the ER was larger than 19dB for both polarizations. A 4 × 4 switch was also demonstrated and the function of dual polarization switching was demonstrated. This approach shows the potential to be applied to the design of polarization-independent on-chip photonic systems with active and passive components such as modulators, cross-waveguides, microrings, etc.